\pdfoutput=1

\documentclass[draft]{ifacconf}

 \usepackage[utf8]{inputenc}
 \usepackage[english]{babel}
 \usepackage[babel]{csquotes}

\usepackage{etex}
\usepackage{amscd}
\usepackage{amsfonts}
\usepackage{amsmath}
\usepackage{amssymb}
\usepackage{color}
\usepackage{array}
\usepackage{tabu}
\usepackage{booktabs}
\usepackage{colortbl}
\usepackage{longtable}
\usepackage{nicefrac}
\usepackage{xspace}
\usepackage{paralist}
\usepackage{listings}
 \usepackage{bashful}
 \usepackage{xifthen}
 \usepackage{url}
 \usepackage{mathpartir}
 \usepackage{versions}
 \usepackage{pdflscape}
 \usepackage{scrtime}
 \usepackage{intcalc}

\usepackage{amsmath}
\newtheorem{theorem}{Theorem}
\newtheorem{example}{Example}
\newtheorem{lemma}{Lemma}
\newtheorem{remark}{Remark}
\newtheorem{proof}{Proof}
\newtheorem{definition}{Definition}
\newtheorem{proposition}{Proposition}
\newtheorem{notation}{Notation}

\usepackage{pgf}
\usepackage{tikz}
\usetikzlibrary{trees,decorations,arrows,automata,shadows,positioning,plotmarks,backgrounds,shapes}
\usetikzlibrary{calc,matrix,fit,petri,decorations.pathmorphing,patterns}
\usetikzlibrary{decorations.pathreplacing}

\tikzstyle{block} = [draw, fill=black!10, rectangle, 
    minimum height=2em, minimum width=3em]
\tikzstyle{blockbig} = [draw, fill=black!10, rectangle, 
    minimum height=2cm, minimum width=2.4cm]
\tikzstyle{sum} = [draw, fill=black!10, circle, node distance=1cm]
\tikzstyle{input} = [coordinate]
\tikzstyle{output} = [coordinate]
\tikzstyle{pinstyle} = [pin edge={to-,thin,black}]
\tikzstyle{sinput} = [coordinate]

\let\theoremOrig\endtheorem
\def\endtheorem{\hspace*{0pt}\hfill\rule{1ex}{1ex}\theoremOrig\vspace*{-1ex}}
\let\exampleOrig\endexample
\def\endexample{\hspace*{0pt}\hfill$\Box$\exampleOrig\vspace*{-1ex}}
\let\lemmaOrig\endlemma
\def\endlemma{\hspace*{0pt}\hfill\rule{1ex}{1ex}\lemmaOrig\vspace*{-1ex}}
\let\remarkOrig\endremark
\def\endremark{\hspace*{0pt}\hfill$\Box$\remarkOrig\vspace*{-1ex}}
\let\proofOrig\endproof
\def\endproof{\hspace*{0pt}\hfill$\Box$\proofOrig\vspace*{-1ex}}
\let\definitionOrig\enddefinition
\def\enddefinition{\hspace*{0pt}\hfill$\Box$\definitionOrig\vspace*{-1ex}}
\let\propositionOrig\endproposition
\def\endproposition{\hspace*{0pt}\hfill\rule{1ex}{1ex}\propositionOrig\vspace*{-1ex}}
\let\notationOrig\endnotation
\def\endnotation{\hspace*{0pt}\hfill$\Box$\notationOrig\vspace*{-1ex}}
\let\corollaryOrig\endcorollary
\def\endcorollary{\hspace*{0pt}\hfill\rule{1ex}{1ex}\corollaryOrig\vspace*{-1ex}}

\usepackage{macros}

\newcommand{\STEP}[1]{\ensuremath{G_{#1}}\xspace}

\usepackage{stmaryrd,pifont}
\newcommand{\ENDOFOUTPUT}{\ensuremath{\diamond}\xspace}

\usepackage{framed}
\makeatletter
\begingroup 
  \makeatletter
  \g@addto@macro\framed{%

  }
\endgroup
\newenvironment{fshaded}{
\MakeFramed {\FrameRestore}}%
{\endMakeFramed}

\makeatletter
\newrobustcmd{\Sven}[1]{\definecolor{shadecolor}{rgb}{.8,.8,1}%
\definecolor{framecolor}{rgb}{0,0,1}%
\begin{fshaded}\textsl{Kommentar Sven: #1}\end{fshaded}}
\newrobustcmd{\Anne}[1]{\definecolor{shadecolor}{rgb}{1,.8,.8}%
\definecolor{framecolor}{rgb}{1,0,0}%
\begin{fshaded}\textsl{Kommentar Anne: #1}\end{fshaded}}
\makeatother

\newrobustcmd{\SINGLEx}[1]{\ensuremath{\underbar{\ensuremath{#1}}}}
\newrobustcmd{\SINGLE}[1]{\List[]{#1}{}{}}

\newrobustcmd{\myFigure}[2]{
\begin{figure}[t]
\footnotesize
\input{figures/#1}
\caption{\normalfont\footnotesize\label{#1}#2}
\end{figure}}
\newrobustcmd{\myFigureWide}[2]{
\begin{figure*}[t]
\footnotesize
\input{figures/#1}
\caption{\normalfont\footnotesize\label{#1}#2}
\end{figure*}}

\newrobustcmd{\myTable}[3]{
\begin{table}[t]
\footnotesize
\input{figures/#1}
\vspace*{2ex}
\caption{\normalfont\footnotesize\label{#2}#3}
\vspace*{-2ex}
\end{table}}
\newrobustcmd{\myTableWide}[3]{
\begin{table*}[t]
\footnotesize
\input{figures/#1}
\caption{\normalfont\footnotesize\label{#2}#3}
\vspace*{-.5cm}
\end{table*}}

\makeatletter
\newlength{\SFS@HEIGHT}
\newlength{\SFS@WIDTH}

\newcommand{\SplitX}[2]{
\settoheight{\SFS@HEIGHT}{$#2$}
\settowidth{\SFS@WIDTH}{$#2$}
\mbox{\begin{tikzpicture}[baseline=(current bounding box.center)]
\node[] (E) at (0,0) {$#1$};
\node[inner sep=0pt] (F) at ($(E.south west)+(1ex,-1ex)+(4ex+.5\SFS@WIDTH,-\SFS@HEIGHT)$) {$#2$};
\node[] (E) at (0,0) {\phantom{$#1$}};
\draw[fill] ($(E.east)+(1ex,0ex)$) circle (.2ex);
\draw[->] ($(E.east)+(1ex,0ex)$) -- ($(E.south east)+(1ex,-1ex)$) -- ($(E.south west)+(1ex,-1ex)$) -- ($(E.south west)+(1ex,-1ex)-(0,\SFS@HEIGHT)$) -- ($(F.west)+(-1ex,0)$);
\end{tikzpicture}}}

\makeatother

\newcommand{\trivialN}[1]{\text{trivial}\xspace}

\newcommand{\FINCOUNTState}[1]{\ensuremath{\#_Q(\ifthenelse{\isempty{#1}}{\cdot}{#1})}}

\newcommand{\UNDEF}{\bot}
\newcommand{\drop}[2]{\mathrm{drop}(#1,#2)}
\newcommand{\length}[1]{|#1|}
\newcommand{\delbot}[1]{\mathrm{del\ENDOFOUTPUT}(#1)}

\newcommand{\ThreeArgCommand}[4][f]{\ensuremath{#1\ifthenelse{\equal{#2#3#4}{}}{}{(\ifthenelse{\equal{#2}{}}{\cdot}{#2},\ifthenelse{\equal{#3}{}}{\cdot}{#3},\ifthenelse{\equal{#4}{}}{\cdot}{#4})}}\xspace}

\newcommand{\GENALGORITHMname}[1][]{\alpha\ifthenelse{\equal{#1}{}}{}{^{#1}}}

\newcommand{\COMP}[4]{\mathsf{Comp}\ifthenelse{\isempty{#1#2#3#4}}{}{(#1,#2,#3,#4)}}

\newcommand{\COMPW}[4]{\mathsf{Comp^W}\ifthenelse{\isempty{#1#2#3#4}}{}{(#1,#2,#3,#4)}}
\newcommand{\COMPUM}[4]{\mathsf{CompUM}\ifthenelse{\isempty{#1#2#3#4}}{}{(#1,#2,#3,#4)}}
\newcommand{\COMPUMW}[4]{\mathsf{CompUM^W}\ifthenelse{\isempty{#1#2#3#4}}{}{(#1,#2,#3,#4)}}

\newcommand{\SIUMpred}[1]{\Phi_{\mathsf{sium}}\ifthenelse{\isempty{#1}}{}{(#1)}}
\newcommand{\UMpred}[1]{\Phi_{\mathsf{um}}\ifthenelse{\isempty{#1}}{}{(#1)}}
\newcommand{\MMpred}[1]{\Phi_{\mathsf{mm}}\ifthenelse{\isempty{#1}}{}{(#1)}}

\newcommand{\ITEMMARKER}{\Yright}
\newcommand{\ITEM}[4]{[#1\fun #2\ITEMMARKER#3,#4]}

\newcommand{\ATS}{\ensuremath{\mathrm{ATS}}\xspace}

\newcommand{\PC}[1]{\overline{#1}}
\newcommand{\PREFIX}[2]{#1 \sqsubseteq #2}
\newcommand{\SUFFIX}[2]{#1 \sqsupseteq #2}

\newcommand{\Parser}{\ensuremath{\mathrm{Parser}}\xspace}

\newcommand{\CFG}{\ensuremath{\mathrm{CFG}}\xspace}
\newcommand{\LR}[1][1]{\ensuremath{\mathrm{LR}(#1)}\xspace}
\newcommand{\DERIVE}[2][]{\ensuremath{\mathrel{\vdash}\ifthenelse{\isempty{#2}}{}{_{#2}}\ifthenelse{\isempty{#1}}{}{^{\ON{#1}}}}\xspace}
\newcommand{\DERIVEst}[2][]{\ensuremath{\mathrel{\vdash^*}\ifthenelse{\isempty{#2}}{}{_{#2}}\ifthenelse{\isempty{#1}}{}{^{\ON{#1}}}}\xspace}
\newcommand{\nDERIVE}[2][]{\ensuremath{\mathrel{\nvdash}\ifthenelse{\isempty{#2}}{}{_{#2}}\ifthenelse{\isempty{#1}}{}{^{\ON{#1}}}}\xspace}
\newcommand{\DERIVATIONS}[3][]{\mathcal{D}\ifthenelse{\isempty{#1}}{}{^{\ON{#1}}}\ifthenelse{\isempty{#2}}{}{_{#2}}\ifthenelse{\isempty{#3}}{}{(#3)}\xspace}

\newcommand{\SENTENTIAL}[3]{\ensuremath{\mathcal{S}\ifthenelse{\isempty{#1}}{}{^{#1}}\ifthenelse{\isempty{#2}}{}{_{#2}}(#3)}\xspace}
\renewcommand{\LANG}[1]{\ensuremath{\ifthenelse{\isempty{#1}}{\mathrm{L_{m}}}{\mathrm{L_{m}}(#1)}}\xspace}
\newcommand{\ULANG}[1]{\ensuremath{\ifthenelse{\isempty{#1}}{\mathrm{L_{um}}}{\mathrm{L_{um}}(#1)}}\xspace}
\newcommand{\LANGc}[1]{\ensuremath{\ifthenelse{\isempty{#1}}{\mathrm{L^c_{m}}}{\mathrm{L^c_{m}}(#1)}}\xspace}
\newcommand{\SCP}{\ensuremath{\mathrm{SCP}}\xspace}

\newcommand{\SDPDA}{\ensuremath{\mathrm{SDPDA}}\xspace}
\newcommand{\NFA}{\ensuremath{\mathrm{NFA}}\xspace}
\newcommand{\DFA}{\ensuremath{\mathrm{DFA}}\xspace}
\newcommand{\EPDA}{\ensuremath{\mathrm{EPDA}}\xspace}
\newcommand{\PDA}{\ensuremath{\mathrm{PDA}}\xspace}
\newcommand{\DPDA}{\ensuremath{\mathrm{DPDA}}\xspace}
\newcommand{\VPTA}{\ensuremath{\mathrm{VPTA}}\xspace}

\newcommand{\EDPDA}{\ensuremath{\mathrm{EDPDA}}\xspace}

\newcommand{\DCFL}[1][]{\ensuremath{\mathrm{DCFL}\ifthenelse{\isempty{#1}}{}{(#1)}}\xspace}
\newcommand{\CFL}[1][]{\ensuremath{\mathrm{CFL}\ifthenelse{\isempty{#1}}{}{(#1)}}\xspace}
\newcommand{\REG}[1][]{\ensuremath{\mathrm{REG}\ifthenelse{\isempty{#1}}{}{(#1)}}\xspace}

\newcommand{\autfont}[1]{#1}

\newcommand{\Aut}[1][]{\autfont{M}^{#1}}
\newcommand{\AutRhs}[1][]{(Q^{#1},\Sigma^{#1},\Gamma^{#1},\delta^{#1},p_0^{#1},\Box^{#1},F^{#1})}

\newcommand{\PCredOff}{\global\edef\PCredIF{X}}\PCredOff
\newcommand{\PCfredPrint}[1]{\ifthenelse{\equal{\PCredIF}{X}}{\begingroup\color{red!50!black}\ON{#1}\endgroup}{\begingroup\color{red}\ON{#1}\endgroup}\PCredOff}
\newcommand{\PCfCC}[3][\SigmaUC]{\PCfredPrint{CC}\ifthenelse{\isempty{#2#3}}{}{(#2,#3,#1)}}
\newcommand{\PCfCCL}[3][\SigmaUC]{\PCfredPrint{CCL}\ifthenelse{\isempty{#2#3}}{}{(#2,#3,#1)}}
\newcommand{\PCfCCS}[3][\SigmaUC]{\PCfredPrint{CCS}\ifthenelse{\isempty{#2#3}}{}{(#2,#3,#1)}}
\newcommand{\PCfAC}[1]{\PCfredPrint{AC}\ifthenelse{\isempty{#1}}{}{(#1)}}
\newcommand{\PCfTE}[2]{\PCfredPrint{TE}\ifthenelse{\isempty{#1#2}}{}{(#1,#2)}\xspace}
\newcommand{\PCfTS}[2]{\PCfredPrint{TS}\ifthenelse{\isempty{#1#2}}{}{(#1,#2)}}
\newcommand{\PCfTRANSIENT}[1]{\PCfredPrint{TRANSIENT}\ifthenelse{\isempty{#1}}{}{(#1)}}
\newcommand{\PCfRNCE}[3]{\PCfredPrint{RNCE}\ifthenelse{\isempty{#1#2#3}}{}{(#1,#2,#3)}}
\newcommand{\PCfSPLIT}[1]{\PCfredPrint{SPLIT}\ifthenelse{\isempty{#1}}{}{(#1)}}
\newcommand{\PCfRULS}[1]{\PCfredPrint{RULS}\ifthenelse{\isempty{#1}}{}{(#1)}}
\newcommand{\PCfCONT}[1]{\PCfredPrint{CONT}\ifthenelse{\isempty{#1}}{}{(#1)}}
\newcommand{\PCfNONBLOCK}[1]{\PCfredPrint{NONBLOCK}\ifthenelse{\isempty{#1}}{}{(#1)}}
\newcommand{\PCfdesc}[3]{\PCfredPrint{desc}\ifthenelse{\isempty{#1#2#3}}{}{(#1,#2,#3)}}
\newcommand{\PCfdescone}[3]{\PCfredPrint{desc1}\ifthenelse{\isempty{#1#2#3}}{}{(#1,#2,#3)}}
\newcommand{\PCfdesconeone}[3]{\PCfredPrint{desc11}\ifthenelse{\isempty{#1#2#3}}{}{(#1,#2,#3)}}
\newcommand{\PCfDollarAugmented}[3]{\PCfredPrint{\$AUG}\ifthenelse{\isempty{#1#2#3}}{}{(#1,#2,#3)}}
\newcommand{\PCfDollarDiminish}[1]{\PCfredPrint{\$DIM}\ifthenelse{\isempty{#1}}{}{(#1)}}
\newcommand{\PCfderive}[3]{\PCfredPrint{derive}\ifthenelse{\isempty{#1#2#3}}{}{(#1,#2,#3)}}
\newcommand{\PCffirst}[2]{\PCfredPrint{first}\ifthenelse{\isempty{#1#2}}{}{(#1,#2)}}
\newcommand{\PCffirstLeqi}[3]{\PCfredPrint{firstleq1}\ifthenelse{\isempty{#1#2#3}}{}{(#1,#2,#3)}}
\newcommand{\PCffirstR}[2]{\PCfredPrint{firstR}\ifthenelse{\isempty{#1#2}}{}{(#1,#2)}}
\newcommand{\PCffirstReduced}[2]{\PCfredPrint{firstRed}\ifthenelse{\isempty{#1#2}}{}{(#1,#2)}}
\newcommand{\PCffirstA}[2]{\PCfredPrint{firstA}\ifthenelse{\isempty{#1#2}}{}{(#1,#2)}}
\newcommand{\PCffirstL}[2]{\PCfredPrint{firstL}\ifthenelse{\isempty{#1#2}}{}{(#1,#2)}}
\newcommand{\PCffirstOne}[2]{\PCfredPrint{first1}\ifthenelse{\isempty{#1#2}}{}{(#1,#2)}}
\newcommand{\PCffirstAll}[2]{\PCfredPrint{firstAll}\ifthenelse{\isempty{#1#2}}{}{(#1,#2)}}
\newcommand{\PCffirstDomStrings}[1]{\PCfredPrint{fDS}\ifthenelse{\isempty{#1}}{}{(#1)}}
\newcommand{\PCfpassesX}[2]{\PCfredPrint{passesX}\ifthenelse{\isempty{#1#2}}{}{(#1,#2)}}
\newcommand{\PCfBASIS}[2]{\PCfredPrint{BASIS}\ifthenelse{\isempty{#1#2}}{}{(#1,#2)}}
\newcommand{\PCfLRPRules}[4]{\PCfredPrint{LRPRules}\ifthenelse{\isempty{#1#2#3#4}}{}{(#1,#2,#3,#4)}}
\newcommand{\PCfLRP}[5]{\PCfredPrint{LRP}\ifthenelse{\isempty{#1#2#3#4#5}}{}{(#1,#2,#3,#4,#5)}}
\newcommand{\PCfREPiip}[1]{\PCfredPrint{REP2+}\ifthenelse{\isempty{#1}}{}{(#1)}}
\newcommand{\PCfREPz}[1]{\PCfredPrint{REP0}\ifthenelse{\isempty{#1}}{}{(#1)}}
\newcommand{\PCfRPP}[1]{\PCfredPrint{RPP}\ifthenelse{\isempty{#1}}{}{(#1)}}
\newcommand{\PCfNDA}[1]{\PCfredPrint{NDA}\ifthenelse{\isempty{#1}}{}{(#1)}}
\newcommand{\PCfLRM}[2]{\PCfredPrint{LRM}\ifthenelse{\isempty{#1#2}}{}{(#1,#2)}}
\newcommand{\PCfLRMonce}[3]{\PCfredPrint{LRM1}\ifthenelse{\isempty{#1#2#3}}{}{(#1,#2,#3)}}
\newcommand{\PCfEPDAGoto}[3]{\PCfredPrint{EPDAGoto}\ifthenelse{\isempty{#1#2#3}}{}{(#1,#2,#3)}}
\newcommand{\PCfEPDAGotoSeq}[3]{\PCfredPrint{EPDAGotoSeq}\ifthenelse{\isempty{#1#2#3}}{}{(#1,#2,#3)}}
\newcommand{\PCfValidEmpty}[2]{\PCfredPrint{ValidEmpty}\ifthenelse{\isempty{#1#2}}{}{(#1,#2)}}
\newcommand{\PCfLRMloop}[5]{\PCfredPrint{LRMloop}\ifthenelse{\isempty{#1#2#3#4#5}}{}{(#1,#2,#3,#4,#5)}}
\newcommand{\PCfGPtoPP}[2]{\PCfredPrint{GP2PP}\ifthenelse{\isempty{#1#2}}{}{(#1,#2)}}
\newcommand{\PCfdescInitial}[1]{\PCfredPrint{descInitial}\ifthenelse{\isempty{#1}}{}{(#1)}}
\newcommand{\PCfSPtoLR}[1]{\PCfredPrint{SP2LR}\ifthenelse{\isempty{#1}}{}{(#1)}}
\newcommand{\PCftimes}[2]{\PCfredPrint{\times}\ifthenelse{\isempty{#1#2}}{}{(#1,#2)}}
\newcommand{\PCfGOTO}[4]{\PCfredPrint{GOTO}\ifthenelse{\isempty{#1#2#3#4}}{}{(#1,#2,#3,#4)}}
\newcommand{\PCfRtoE}[2]{\PCfredPrint{R2E}\ifthenelse{\isempty{#1#2}}{}{(#1,#2)}}
\newcommand{\PCfRtoQ}[2]{\PCfredPrint{R2Q}\ifthenelse{\isempty{#1#2}}{}{(#1,#2)}}
\newcommand{\PCfRENAMEQ}[2]{\PCfredPrint{RenQ}\ifthenelse{\isempty{#1#2}}{}{(#1,#2)}}
\newcommand{\PCfRENAMEG}[2]{\PCfredPrint{RenG}\ifthenelse{\isempty{#1#2}}{}{(#1,#2)}}
\newcommand{\PCfFILTER}[2]{\PCfredPrint{filter}\ifthenelse{\isempty{#1#2}}{}{(#1,#2)}}
\newcommand{\PCfFILTERfirst}[2]{\PCfredPrint{filter1}\ifthenelse{\isempty{#1#2}}{}{(#1,#2)}}
\newcommand{\PCfPostCl}[1]{\PCfredPrint{PostCl}\ifthenelse{\isempty{#1}}{}{(#1)}}
\newcommand{\PCfREPLACE}[3]{\PCfredPrint{replace}\ifthenelse{\isempty{#1#2#3}}{}{(#1,#2,#3)}}
\newcommand{\PCfRUP}[1]{\PCfredPrint{RUP}\ifthenelse{\isempty{#1}}{}{(#1)}}
\newcommand{\PCfRNPP}[1]{\PCfredPrint{RNPP}\ifthenelse{\isempty{#1}}{}{(#1)}}
\newcommand{\PCfRNPPL}[2]{\PCfredPrint{RNPPL}\ifthenelse{\isempty{#1#2}}{}{(#1,#2)}}
\newcommand{\PCfRNPPone}[2]{\PCfredPrint{RNPP1}\ifthenelse{\isempty{#1#2}}{}{(#1,#2)}}
\newcommand{\PCfRUS}[1]{\PCfredPrint{RUS}\ifthenelse{\isempty{#1}}{}{(#1)}}
\newcommand{\PCfFreshState}[2]{\PCfredPrint{FreshState}\ifthenelse{\isempty{#1#2}}{}{(#1,#2)}\xspace}
\newcommand{\PCfRNPS}[1]{\PCfredPrint{RNPS}\ifthenelse{\isempty{#1}}{}{(#1)}}

\newcommand{\SigmaUC}{\Sigma_{\mathsf{uc}}}

\newcommand{\NoAcTr}[1]{\ON{NoAcTr}\ifthenelse{\isempty{#1}}{}{(#1)}}
\newcommand{\NoAcSt}[1]{\ON{NoAcSt}\ifthenelse{\isempty{#1}}{}{(#1)}}

\newcommand{\REMOVEUSELESSSTATES}[1]{\ON{RULS}\ifthenelse{\isempty{#1}}{}{(#1)}}

\newcommand{\RCP}[1]{\ON{RemoveContProblems}\ifthenelse{\isempty{#1}}{}{(#1)}}
\newcommand{\NONBLOCK}[1]{\ON{EnsureNonBlocking}\ifthenelse{\isempty{#1}}{}{(#1)}}

\newcommand{\CC}[2]{\ON{CC}\ifthenelse{\isempty{#1#2}}{}{(#1,#2)}}
\newlength{\XInnerSep}
\usetikzlibrary{calc, intersections,positioning}
\usepackage{graphicx,adjustbox}
\usepackage{scrtime}
\usepackage{natbib}
\newcommand{\TOBEREMOVED}[2][1]{{\ifthenelse{\equal{#1}{1}}{\color{blue!80!white}#2}{\color{blue!50!white}#2}}}
\newcommand{\Derivations}[1]{\mathcal{D}(#1)\xspace}
\newcommand{\InitialDerivations}[1]{\mathcal{D}_\mathrm{I}(#1)\xspace}
\newcommand{\AccLang}[1]{\ON{L_{m}}(#1)\xspace}

\newcommand{\AnyLang}[1]{\ON{L_{um}}(#1)\xspace}

\newcommand{\IFTHENELSE}[3]{\ON{if }#1\ON{ then }#2\ON{ else }#3\xspace}
\newcommand{\myparagraph}[1]{\par$\blacktriangleright$\emph{#1:}}

\newcommand{\AnConf}[2]{\mathrm{\maltese}(#1,#2)\xspace}

\newcommand{\CConf}[1]{\mathcal{C}(#1)\xspace}

\newcommand{\ReachConf}[1]{\mathcal{C}_{\ON{reach}}(#1)}

\newcommand{\OUTm}[1]{o_{\var{m}}\ifthenelse{\isempty{#1}}{}{(#1)}}
\newcommand{\OUTum}[1]{o_{\var{um}}\ifthenelse{\isempty{#1}}{}{(#1)}}

\newcommand{\ATSrhsX}{E,C,S,\pi_S,R,c_0,A,O,\OUTm{},\OUTum{}}
\newcommand{\ATSrhs}{(\ATSrhsX)}

\newcommand{\kPrefix}[1]{{1}{:}#1\xspace}
\newcommand{\kPREFIX}[2][k]{{#1}{:}#2\xspace}

\newcommand{\RULE}[4]{#1 {\mid} #2 {\fun} #3 {\mid} #4}

\newcommand{\OBF}{operational blockfree\xspace}

\newrobustcmd{\FIGREF}[1]{Figure~\ref{#1}}
\newrobustcmd{\THMREF}[1]{Theorem~\ref{#1}}
\newrobustcmd{\DEFREF}[1]{Definition~\ref{#1}}
\newrobustcmd{\TABREF}[1]{Table~\ref{#1}}
\newrobustcmd{\SECREF}[1]{Section~\ref{#1}}
\newrobustcmd{\EQNREF}[3]{Equation\ifthenelse{\isempty{#2#3}}{}{s}~\eqref{#1}\ifthenelse{\isempty{#2#3}}{}{#2\eqref{#3}}}

\newcommand{\instantiate}[1]{{\fboxsep1pt\fbox{$\vphantom{\ATSrhsX}#1$}}}

\begin{document}
\def\abstractname{{\bfseries {Abstract: }}}
\def\figurename{\bfseries\footnotesize Figure}
\def\tablename{\bfseries\footnotesize Table}
\begin{frontmatter}
\title{Enforcing Operational Properties\\including Blockfreeness for\\Deterministic Pushdown Automata}

\author{S. Schneider and U. Nestmann}

\address{Technische Universität Berlin}
\begin{abstract}
We present an algorithm which modifies a deterministic pushdown automaton (\DPDA) such that
\begin{inparaenum}[(i)]\item the marked language is preserved, \item lifelocks are removed, \item deadlocks are removed, \item all states and edges are accessible, and \item \OBF{}ness is established (i.e., coaccessibility in the sense that every initial derivation can be continued to a marking configuration). \end{inparaenum}
This problem can be trivially solved for deterministic finite automata (\DFA) but is not solvable for standard petri net classes.
The algorithm is required for an operational extension of the supervisory control problem (\SCP) to the situation where the specification in modeled by a \DPDA.
\end{abstract}


\end{frontmatter}

We are introducing an algorithm to transform a \DPDA such that its observable operational behavior is restricted to its desired fragment.
The algorithm decomposes the problem into three steps: transformation of the \DPDA into a Context Free Grammar (\CFG) while preserving the operational behavior, restricting the \CFG to enforce operational blockfreeness, and the transformation of the resulting \CFG via \Parser{}s to \DPDA while preserving and establishing the relevant criteria on the operational behavior.
The algorithm presented here is an essential part for the effective solution of the supervisory control problem for \DFA plants and \DPDA specifications which is reduced (in the companion paper by \citet*{SchneiderSchmuck2014}) to the effective implementability of ensuring blockfreeness (solved in this paper) and ensuring controllability (solved in the companion paper by \citet*{SchmuckSchneider2014}).

In \SECREF{sect:ATS} we define abstract transition systems (\ATS) as a basis for the systems involved in the algorithm and give a formal problem statement to be solved for \DPDA.
In \SECREF{sect:CTS} we define the concrete transition systems appearing in the algorithm as instantiations of \ATS.
In \SECREF{sect:ALG} we present the extensive algorithm due to space restrictions mostly informally using a running example before we discuss the formal verification and possible improvements of the approach.
The formal constructions of the algorithm are contained in~\cite{SchneiderSchmuck_TechRep_2013}.
We summarize our results in \SECREF{sect:SUM} and outline our next steps in \SECREF{sect:FUTURE}.

\section{Abstract Transition Systems}\label{sect:ATS}
The concrete systems used in this paper (including \DPDA, \CFG, and \Parser{}s) are instantiations of the subsequently defined class of Abstract Transition Systems (\ATS).
Thus, they will inherit the uniform definitions of derivations, languages, and the problem to be solved from the \ATS definitions.

\newrobustcmd{\RED}[1]{{\color{red}#1}}
\newrobustcmd{\VISIBLE}[1]{#1}
\newrobustcmd{\OPTSIGMAELEM}{\VISIBLE{\sigma}}
\newrobustcmd{\STACKELEM}{\VISIBLE{\gamma}}
\newrobustcmd{\SIGMAELEM}{\VISIBLE{\alpha}}
\newrobustcmd{\STACKSTR}{\VISIBLE{s}}
\newrobustcmd{\SIGMASTR}{\VISIBLE{w}}
\newrobustcmd{\FIXEDSTR}{\VISIBLE{f}}
\newrobustcmd{\STATEELEM}{\VISIBLE{p}}
\newrobustcmd{\STATESTR}{\VISIBLE{\tilde{p}}}
\newrobustcmd{\NONTERMELEMA}{\VISIBLE{A}}
\newrobustcmd{\NONTERMELEMB}{\VISIBLE{B}}
\newrobustcmd{\CFGMIXSTR}{\VISIBLE{v}}
\newrobustcmd{\CFGMIXELEM}{\VISIBLE{\kappa}}

Throughout the paper we use the following notations.
\begin{notation}
Let $A$ be an alphabet and let $B$ be a set. Then
\begin{inparaenum}[(i)]
\item $A^*$ denotes the set of all finite words over $A$,
\item $A^{\leq1}=A\cup\Set{\EMPTYSTRING}$,
\item $A^{\omega *}$ denotes the set of all finite and infinite words over $A$,
\item $\sconc$ is the (sometimes omitted) concatenation operation on words (and languages), 
\item $\PREFIX{}{}$ is the prefix relation, 
\item $\PC{A}$ is the prefix-closure of $A$, 
\item $\SUFFIX{}{}$ is the suffix relation, and 
\item $\kPREFIX{\SIGMASTR}$ denotes the k-Prefix of $\SIGMASTR\in A^*$ defined by $(\IFTHENELSE{\propConj{\SIGMASTR=\SINGLE{\SIGMAELEM}\sconc \SIGMASTR'}{k>0}}{\SINGLE{\SIGMAELEM}\sconc(\kPREFIX[(k-1)]{\SIGMASTR'})}{\EMPTYSTRING})$, and
\item $\AnConf{A}{B}$ denotes $\CARTPROD{\UNION*{A}{\Set{\UNDEF}}}{B}$ where $\UNDEF$ represents undefinedness.
\end{inparaenum}
\end{notation}

\begin{definition}[Abstract Transition System]\label{def:ATS}\leavevmode\\
$\mathcal{S}=\ATSrhs\in\ATS$ \IFFtext
\begin{inparaenum}[(i)]
\item $E$ is a set of step-edges,
\item $C$ is a set of configurations,
\item $S$ is a set of states,
\item $\pi_S$ maps each configuration to at most one state,
\item $R$ is a binary step-relation on $\AnConf{E}{C}$,
\item $c_0\in C$ is the initial configuration,
\item $A$ is the marking subset of $C$,
\item $O$ is the set of outputs, and
\item $\OUTum{}:C\fun \POWERSET{O}$ and $\OUTm{}:A\fun \POWERSET{O}$ define the unmarked and marked outputs for configurations.
\end{inparaenum}
\end{definition}
For these \ATS we define their derivations, generated languages, and subsequently the properties to be enforced.
\begin{definition}[Semantics of \ATS]
\begin{inparaenum}[(i)]
\item the set of derivations $\Derivations{\mathcal{S}}$ contains all elements from $\AnConf{E}{C}^{\omega *}$ starting in a configuration of the form $(\UNDEF,c)$ where all adjacent $(c_1,e_1),(c_2,e_2)\in\AnConf{E}{C}$ satisfy $(c_1,e_1)\mathrel{R}(c_2,e_2)$,
\item the set of initial derivations $\InitialDerivations{\mathcal{S}}$ contains all elements of $\Derivations{\mathcal{S}}$ starting with $(\UNDEF,c_0)$,
\item the reachable configurations $\ReachConf{\mathcal{S}}$ are defined by $\SetComp{c\in C}{\ExQ{d\in\InitialDerivations{\mathcal{S}}}{d(n)=(e,c)}}$,
%
\item the marked language $\AccLang{\mathcal{S}}$ is defined by $\cup \OUTm{\INTERSECT{F}{\ReachConf{\mathcal{S}}}}$, and
\item the unmarked language $\AnyLang{\mathcal{S}}$ is defined by $\cup \OUTum{\ReachConf{\mathcal{S}}}$.
\end{inparaenum}\\
The concatenation of derivations $d_1,d_2\in\Derivations{\mathcal{S}}$ is given by $(d_1\sconc_nd_2)(i)=(\IFTHENELSE{i\le n}{d_1(i)}{d_2(i-n)})$.
\end{definition}
\begin{definition}[Properties of \ATS]\label{ATS:props}
\begin{inparaenum}[(i)]
\item $\mathcal{S}$ has a deadlock \IFFtext for some finite $d\in\InitialDerivations{\mathcal{S}}$ of length $n\in\NAT$ which is not marking (i.e., for all $k$, $d(k)=(e,c)$ implies $c\notin A$) there is no $c'$ such that $d(n)\mathrel{R}c'$,
\item $\mathcal{S}$ has a lifelock \IFF for some infinite $d\in\InitialDerivations{\mathcal{S}}$ there is an $N\in\NAT$ such that the unmarked language of $d$ is constant after $N$ (i.e., for all $k\geq N$, $\OUTum{d(N)}=\OUTum{d(k)}$),
\item $\mathcal{S}$ is accessible \IFF for each $\STATEELEM\in S$ there is $c\in\ReachConf{\mathcal{S}}$ such that $\pi_S(c)=\STATEELEM$ and
for each $e\in E$ there is $d\in\InitialDerivations{\mathcal{S}}$ such that $d(n)=(e,c)$, and
\item $\mathcal{S}$ is \OBF \IFF for any finite $d_i\in\InitialDerivations{\mathcal{S}}$ of length $n\in\NAT$ ending in $d_i(n)=(e,c)$ there is a continuation $d_c\in\Derivations{\mathcal{S}}$ such that $d_i\sconc_n d_c$ is a marking derivation and $d_i$ and $d_c$ match at the gluing point $n$ (i.e., $d_c(0)=(\UNDEF,c)$).
\end{inparaenum}
\end{definition}

By definition, for \OBF \ATS the absence of deadlocks is guaranteed.
Finally, we present the problem of enforcing the desired properties on an \ATS, which will be solved for \DPDA by the algorithm presented in \SECREF{sect:ALG}.

\begin{definition}[Problem Statement for \ATS]\label{def:ATS:problem}
Let $\mathcal{S}\in\ATS$.
How to find $\mathcal{S}'\in\ATS$ such that 
\begin{inparaenum}[(i)]
\item $\AccLang{\mathcal{S}}=\AccLang{\mathcal{S}'}$,
\item $\mathcal{S}'$ is accessible,
\item $\mathcal{S}'$ has no deadlocks,
\item $\mathcal{S}'$ has no lifelocks, and
\item $\mathcal{S}'$ is \OBF?
\end{inparaenum}
\end{definition}
In the \DFA-setting: lifelocks can not occur and the other aspects of the problem are solved by simple and efficient graph-traversal algorithms pruning out states which are either not reachable from the initial state or from which no marking state can be reached\footnote{The trivial handling of an \ATS with empty marked language obtained at some point of the calculation is kept implicit in this paper (in this case, no solution exists and the calculation can be aborted).}.

\section{Concrete Transition Systems}\label{sect:CTS}
Every deterministic context free language can be properly represented by at least three different types of finite models: a deterministic \EPDA, a context free grammar (\CFG) satisfying the \LR determinism property, and a deterministic \Parser.
These three types occur at intermediate steps of our algorithm which solves the problem stated in \DEFREF{def:ATS:problem}.
Therefore, the following subsections contain their definitions as instantiations of the \ATS.
In each of the three cases we proceed in three steps:
\begin{inparaenum}
\item definition of \EPDA, \CFG, and \Parser as tuples,
\item instantiation of the \ATS-scheme by defining each of the ten components, and
\item characterization of the determinism conditions.
\end{inparaenum}

\begin{remark}
We provide the slightly nonstandard branching semantics\footnote{The branching interpretation is already the standard for \CFG.} for \EPDA and \Parser{}s which utilize a history variable in the configurations to greatly simplify the definition of the operational-blockfreeness from \DEFREF{ATS:props}.
Furthermore, this branching semantics corresponds to the intuition that the finite state realizations are generators rather than acceptors of languages, as it is customary in the context of supervisory control theory.

\end{remark}
\newrobustcmd{\ENDOFSTACK}{\ensuremath{\Box}\xspace}
\subsection{\EPDA and \DPDA}
We introduce \EPDA, which are \NFA enriched with a variable on which the stack-operations top, pop, and, push can be executed.
\begin{definition}[Extended Pushdown Automata (\EPDA)]\label{def:PDA}\leavevmode\\
$\Aut=\AutRhs\in\EPDA$ \IFF
\begin{inparaenum}[(i)]
\item the states $Q$, the output alphabet $\Sigma$, the stack alphabet $\Gamma$, and the set of edges $\delta$ are finite ($Q$, $Q^*$, $\Sigma$, $\Sigma^*$, $\Gamma$, $\Gamma^*$ range over $\STATEELEM$, $\STATESTR$, $\SIGMAELEM$, $\SIGMASTR$, $\STACKELEM$, $\STACKSTR$, respectively),
\item $\delta:\CARTPROD{Q}{\CARTPROD{\CARTPROD{\CARTPROD{\Sigma^{\leq1}}{\Gamma^*}}{\Gamma^*}}}{Q}$,
\item the end-of-stack marker $\ENDOFSTACK$ is contained in $\Gamma$,
\item the marking states $F$ and the initial state $\STATEELEM_0$ are contained in $Q$, and
\item $\ENDOFSTACK$ is never removed from the stack (i.e., $(\STATEELEM,\OPTSIGMAELEM,\STACKSTR,\STACKSTR',\STATEELEM')\in\delta$ and $\SUFFIX{\STACKSTR}{\ENDOFSTACK}$ imply $\SUFFIX{\STACKSTR'}{\ENDOFSTACK}$).
\end{inparaenum}
\end{definition}
We proceed with the \ATS instantiation for \EPDA.
\begin{definition}[\EPDA---\ATS Instantiation]\label{def:PDAsemantics}\leavevmode\\
An \EPDA $\Aut=\AutRhs$ instantiates the \ATS scheme $\ATSrhs$ via:
\begin{inparaenum}[(i)]
\item \instantiate{E}~$\delta$
\item \instantiate{C}~$\CConf{\Aut}\deff\CARTPROD{Q}{\CARTPROD{\Sigma^*}{\Gamma^*}}$ where $(\STATEELEM,\SIGMASTR,\STACKSTR)\in\CConf{\Aut}$ consists of a state $\STATEELEM$, a history variable $\SIGMASTR$ (storing the symbols generated), and the stack-variable $\STACKSTR$
\item \instantiate{S}~$Q$
\item \instantiate{\pi_S(\STATEELEM,\SIGMASTR,\STACKSTR)}~$\STATEELEM$
\item \instantiate{R}~$\DERIVE{\Aut}:\AnConf{\delta}{\CConf{\Aut}}^2$ defined by $(e,(\STATEELEM,\SIGMASTR,\STACKSTR'\sconc \STACKSTR))\DERIVE{\Aut}((\STATEELEM,\OPTSIGMAELEM,\STACKSTR',\STACKSTR'',\STATEELEM'),\STATEELEM',\SIGMASTR\sconc \OPTSIGMAELEM,\STACKSTR''\sconc \STACKSTR)$
\item \instantiate{c_0} $(\STATEELEM_0,\EMPTYSTRING,\ENDOFSTACK)$
\item \instantiate{A}~$\SetComp{(\STATEELEM,\SIGMASTR,\STACKSTR)\in\CConf{\Aut}}{\STATEELEM\in F}$
\item \instantiate{O}~$\Sigma^*$
\item \instantiate{\OUTm{\STATEELEM,\SIGMASTR,\STACKSTR},\OUTum{\STATEELEM,\SIGMASTR,\STACKSTR}}~$\Set{\SIGMASTR}$
\end{inparaenum}
\end{definition}

\newcommand{\Yes}{\checkmark}

The well known sub-classes of \EPDA having one or more of the properties below are defined in \TABREF{tab:subclasses}.
\begin{definition}[Sub-classes of \EPDA]
\myTable{fig__subclasses}{tab:subclasses}{Subclasses of \EPDA.}
An \EPDA is 1-pop\-ping \IFFtext every edge pops precisely one element from $\Gamma$ from the stack.
An \EPDA is deterministic \IFFtext for every reachable configuration all two distinct steps append distinct elements of $\Sigma$ to the history variable\footnote{Thus, $\EMPTYSTRING$-steps may not be enabled simultaneously with other steps.}.
An \EPDA is $\EMPTYSTRING$-step-free \IFFtext no edge is of the form $(\STATEELEM,\EMPTYSTRING,\STACKSTR,\STACKSTR',\STATEELEM')$.
An \EPDA is stack-free \IFFtext every edge is of the form $(\STATEELEM,\SIGMAELEM,\SINGLE{\ENDOFSTACK},\SINGLE{\ENDOFSTACK},\STATEELEM')$.
\end{definition}

\subsection{\CFG and \LR}
A \CFG (e.g., defined by \cite{DBLP:journals/iandc/GinsburgG66b}) is a term-replacement system replacing a nonterminal with a word over output symbols\footnote{The output symbols of a \CFG are usually called terminals.} and nonterminals.
\begin{definition}[Context-Free Grammars (\CFG)]\label{def:cfg}\leavevmode\\
$G=\Tuple{N,\Sigma,P,S}\in\CFG$ \IFF
\begin{inparaenum}[(i)]
\item the nonterminals $N$ (ranging over $\NONTERMELEMA,\NONTERMELEMB$), the output alphabet $\Sigma$, and the productions $P$ are finite
\item $P:\CARTPROD{N}{\UNION*{N}{\Sigma}^*}$, and
\item the initial nonterminal $S$ is contained in $N$.
\end{inparaenum}
$\UNION{N}{\Sigma}$ and $\UNION*{N}{\Sigma}^*$ range over $\CFGMIXELEM$ and $\CFGMIXSTR$, respectively.
Productions $(\NONTERMELEMA,\CFGMIXSTR)$ are written $\NONTERMELEMA\fun\CFGMIXSTR$.
\end{definition}
\begin{definition}[\CFG---\ATS Instantiation]\label{def:CFGsemantics}\leavevmode\\
A $\CFG$ $G=\Tuple{N,\Sigma,P,S}$ instantiates the \ATS scheme $\ATSrhs$ via:
\begin{inparaenum}[(i)]
\item \instantiate{E}~$P$
\item \instantiate{C}~$\CConf{G}$ $=\UNION*{N}{\Sigma}^*$
\item \instantiate{S}~$N$
\item \instantiate{\pi_S}~take the first nonterminal (if present) of the configuration
\item \instantiate{R}~$\DERIVE{G}:\AnConf{P}{\CConf{G}}^2$ given by $(e,(\CFGMIXSTR_1\sconc\NONTERMELEMA\sconc \CFGMIXSTR_2))\DERIVE{G}((\NONTERMELEMA,\CFGMIXSTR),\CFGMIXSTR_1\sconc\CFGMIXSTR\sconc\CFGMIXSTR_2)$
\item \instantiate{c_0}~$S$
\item \instantiate{A}~$\Sigma^*$
\item \instantiate{O}~$\Sigma^*$
\item \instantiate{\OUTm{\CFGMIXSTR}}~$\Set{\CFGMIXSTR}$
\item \instantiate{\OUTum{\CFGMIXSTR}}~$\PC{\Set{\CFGMIXSTR}}$
\end{inparaenum}
\end{definition}
The \LR-condition below, which corresponds to the determinism property of \EPDA, depends on the restriction of the step-relation to the replacement of the right-most nonterminal which will be denoted by the index $\mathrm{rm}$.
\begin{definition}[\LR-Condition]
According to \cite{sippu-2} (page 52)\footnote{The here relevant section 6.6 of the monograph \cite{sippu-2} is based primarily on the work of \cite{DBLP:journals/iandc/Knuth65} which was later extended by \cite{Aho:1972:TPT:578789}.}, \LR is the set of all \CFG for which (assuming $x\in\Sigma^*$)\\
\begin{inparaenum}[(i)]
\item $(\UNDEF,S)\DERIVE[rm*]{G}(e_1,\CFGMIXSTR_1'\sconc \NONTERMELEMA_1\sconc \SIGMASTR_1)\DERIVE[rm]{G}((\NONTERMELEMA_1,\CFGMIXSTR_1),\CFGMIXSTR_1'\sconc\CFGMIXSTR_1\sconc \SIGMASTR_1)$,\\
\item $(\UNDEF,S)\DERIVE[rm*]{G}(e_2,\CFGMIXSTR_2'\sconc \NONTERMELEMA_2\sconc \SIGMASTR_2)\DERIVE[rm]{G}((\NONTERMELEMA_2,\CFGMIXSTR_2),\CFGMIXSTR_2'\sconc\CFGMIXSTR_2\sconc \SIGMASTR_2)$,\\
\item $\CFGMIXSTR_2'\sconc\CFGMIXSTR_2=\CFGMIXSTR_1'\sconc\CFGMIXSTR_1\sconc x$, and\\
\item $\kPrefix{\SIGMASTR_1}=\kPrefix{(x\sconc \SIGMASTR_2)}$, imply\\
\item $\CFGMIXSTR_1'=\CFGMIXSTR_2'$, $\NONTERMELEMA_1=\NONTERMELEMA_2$, and $\CFGMIXSTR_1=\CFGMIXSTR_2$.
\end{inparaenum}
\end{definition}
Intuitively, if a parser for a \CFG has generated the shorter prefix $\CFGMIXSTR_1'\sconc\CFGMIXSTR_1$ it must be able to decide by fixing the next symbol ($\kPrefix{\SIGMASTR_1}$ and $\kPrefix{(x\sconc \SIGMASTR_2)}$, respectively) whether $(\NONTERMELEMA_1,\CFGMIXSTR_1)$ is to be applied backwards or whether for $x\neq\EMPTYSTRING$ another symbol of $x$ should be generated or for $x=\EMPTYSTRING$ the production $(\NONTERMELEMA_2,\CFGMIXSTR_2)$ is to applied backwards\footnote{E.g., $\Tuple{\Set{S,A,B},\Set{a},\Set{\Tuple{S,A},\Tuple{S,B},\Tuple{A,a},\Tuple{B,a}},S}\notin\LR$.}.

\subsection{\Parser}
Intuitively, a \Parser is an \EPDA with mild modifications\footnote{An equivalent linear/scheduled definition of \Parser is given by \cite{sippu-2}.}:
\begin{inparaenum}
\item the parser may fix the next output-symbol (without generating it) and
\item the parser may terminate the generation of symbols (by fixing the end-of-output marker $\ENDOFOUTPUT$).
\end{inparaenum}
\begin{definition}[\Parser]\label{def:parser}
$M=\Tuple{N,\Sigma,S,F,P,\ENDOFOUTPUT}\in\Parser$ \IFF\\
\begin{inparaenum}[(i)]
\item the stack alphabet $N$, the output alphabet $\Sigma$, the marking stack-tops $F$, and the rules $P$ are finite,
($N$, $N^*$, $\Sigma$, $\Sigma^*$, range over $\STATEELEM$, $\STATESTR$, $\SIGMAELEM$, $\SIGMASTR$, respectively)
\item $P:\CARTPROD{\CARTPROD{N^+}{\Sigma^{\le1}}}{\CARTPROD{N^+}{\Sigma^{\le1}}}$,
\item the initial stack symbol $S$ and the marking stack-tops $F$ are contained in $N$,
\item the end-of-output marker $\ENDOFOUTPUT$ is contained in $\Sigma$,
\item the parser may not modify the output (i.e., $\Tuple{\STACKSTR\sconc\STATEELEM,\SIGMASTR,\STACKSTR'\sconc\STATEELEM',\SIGMASTR'}\in P$ implies $\SUFFIX{\SIGMASTR}{\SIGMASTR'}$ (i.e., $\SIGMASTR$ ends with $\SIGMASTR'$)), and
\item the end-of-output marker $\ENDOFOUTPUT$ may not be generated (i.e., $\Tuple{\STACKSTR\sconc\STATEELEM,\SIGMASTR\sconc\SIGMASTR',\STACKSTR'\sconc\STATEELEM',\SIGMASTR'}\in P$ and $\SUFFIX{\SIGMASTR}{\ENDOFOUTPUT}$ imply $\SUFFIX{\SIGMASTR'}{\ENDOFOUTPUT}$).
\end{inparaenum}
Rules $(\STACKSTR\sconc\STATEELEM,\SIGMASTR\sconc\SIGMASTR',\STACKSTR'\sconc\STATEELEM',\SIGMASTR')$ are written $\RULE{\STACKSTR\sconc\STATEELEM}{\SIGMASTR\sconc\SIGMASTR'}{\STACKSTR'\sconc\STATEELEM'}{\SIGMASTR'}$.
\end{definition}
Intuitively, a rule $\RULE{\STACKSTR\sconc\STATEELEM}{\SIGMASTR\sconc\SIGMASTR'}{\STACKSTR'\sconc\STATEELEM'}{\SIGMASTR'}$ is changing the state from $\STATEELEM$ to $\STATEELEM'$, pops $\STACKSTR$ from the stack, pushes $\STACKSTR'$ to the stack, fixes the output $\SIGMASTR'$, and generates $\SIGMASTR$ to the output.
\begin{definition}[\Parser---\ATS Instantiation]\label{def:ParserSemantics}\leavevmode\\
A \Parser $M=\Tuple{N,\Sigma,S,F,P,\ENDOFOUTPUT}$ in\-stan\-ti\-ates the \ATS scheme
 $\ATSrhs$ via:
\begin{inparaenum}[(i)]
\item \instantiate{E}~$P$
\item \instantiate{C}~$\CConf{M}\deff\CARTPROD{N^+}{\CARTPROD{\Sigma^*}{\Sigma^*}}$ where $(\STACKSTR\sconc \STATEELEM,\SIGMASTR,\FIXEDSTR)\in\CConf{M}$ contains the stack fragment $\STACKSTR$, the current state $\STATEELEM$, a history variable $\SIGMASTR$, and the \emph{fixed} part $\FIXEDSTR\in\Sigma^{\leq1}$ which the parser fixed without generating it.
\item \instantiate{S}~$\SetComp{\overline{\STATEELEM}\in N}{\propConj{(\STACKSTR\sconc\STATEELEM,\SIGMASTR\sconc\SIGMASTR',\STACKSTR'\sconc\STATEELEM',\SIGMASTR')\in P}{\overline{\STATEELEM}\in\Set{\STATEELEM,\STATEELEM'}}}$
\item \instantiate{\pi_S(\STACKSTR\sconc \STATEELEM,\SIGMASTR,\FIXEDSTR)}~$\Set{\STATEELEM}$
\item \instantiate{R}~$\DERIVE{M}:\AnConf{P}{\CConf{M}}^2$ given by $(e,(\STACKSTR\sconc\STACKSTR_1\sconc\STATEELEM,\SIGMASTR,\FIXEDSTR))\DERIVE{M}((\STACKSTR_1\sconc\STATEELEM,\SIGMASTR_1,\STACKSTR_2\sconc\STATEELEM',\SIGMASTR_2),\STACKSTR\sconc\STACKSTR_2\sconc\STATEELEM',\SIGMASTR',\FIXEDSTR')$ where
  \begin{inparaenum}[(a)]
  \item $\propDisj{\PREFIX{\SIGMASTR_1}{\FIXEDSTR}}{\PREFIX{\FIXEDSTR}{\SIGMASTR_1}}$,
  \item $\SIGMASTR'=\SIGMASTR\sconc \drop{\length{\FIXEDSTR}}{\delbot{\SIGMASTR_1}}$,\!\footnote{Here $\delbot{\overline{\SIGMASTR}}$ removes a potential $\ENDOFOUTPUT$ from the end of $\overline{\SIGMASTR}$ and $\drop{n}{\overline{\SIGMASTR}}$ drops the first $n$ symbols from $\overline{\SIGMASTR}$.} and finally
  \item $\FIXEDSTR'=\SIGMASTR_2\sconc \drop{\length{\SIGMASTR_1}}{\FIXEDSTR}$.
  \end{inparaenum}
\item \instantiate{c_0}~$(S,\EMPTYSTRING,\EMPTYSTRING)$
\item \instantiate{A}~$\SetComp{(\STACKSTR\sconc \STATEELEM,\SIGMASTR,\FIXEDSTR)}{\propConj{\FIXEDSTR\in\Set{\EMPTYSTRING,\ENDOFOUTPUT}}{\STATEELEM\in F}}$
\item \instantiate{O}~$\Sigma^*$
\item \instantiate{\OUTm{\STACKSTR\sconc \STATEELEM,\SIGMASTR,\FIXEDSTR}}~$\Set{\SIGMASTR}$
\item \instantiate{\OUTum{\STACKSTR\sconc \STATEELEM,\SIGMASTR,\FIXEDSTR}}~$\Set{\SIGMASTR}$.
\end{inparaenum}
\end{definition}
A \Parser is deterministic \IFF for all reachable configuration all two distinct steps
\begin{inparaenum}[(i)]
\item append distinct symbols to the history variable, or
\item one step adds a symbol to the history variable and the other step completes the output-generation by fixing the end-of-output marker $\ENDOFOUTPUT$\footnote{A parser may (depending on the other rules) be deterministic if $(\STATEELEM_1,w\sconc \SIGMAELEM,\EMPTYSTRING)$ and $(\STATEELEM_2,w,\ENDOFOUTPUT)$ are successors of the same reachable configuration $(p,w,\EMPTYSTRING)$.}.
\end{inparaenum}

\section{Approach}\label{sect:ALG}
\newcommand{\EdgeLabel}[3]{{#1}\kern-1pt{,}\kern-1pt{#2}\kern-1pt{,}\kern-1pt{#3}}
\newcommand{\EdgeSep}{;}

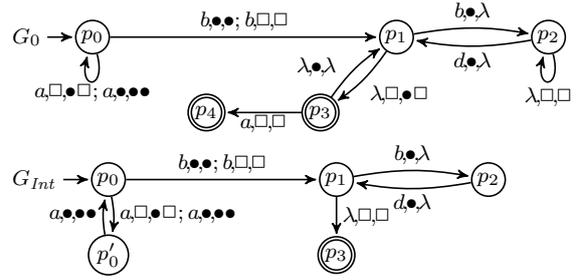
\myFigure{fig__example__M_0_and_M_I}{\DPDA \STEP{0} and $G_{\mathit{Int}}$ generating $\SetComp{a^{2n}b(bd)^n}{n\in\NAT}$.}
\emph{Motivation:} For example, the \DPDA \STEP{0} in \FIGREF{fig__example__M_0_and_M_I} exhibits
\begin{inparaenum}[]
\item a lifelock generating the output $b$ reaching $\STATEELEM_1,\STATEELEM_3$ arbitrarily often,
\item a lifelock (and blocking situations) generating the output $abb$ reaching $\STATEELEM_2$ arbitrarily often,
\item a non accessible state $\STATEELEM_4$ (along with the edge leading to it), but
\item no deadlock.
\end{inparaenum}
Observe that the cause ($\STEP{0}$ does not properly distinguish between an even or odd number of generated $a$s) is structurally separated from the lifelock at $\STATEELEM_2$.
Thus, the intuitive solution $G_{\mathit{Int}}$ (see \FIGREF{fig__example__M_0_and_M_I}) is obtained by splitting the state $\STATEELEM_0$ and by removing junk.
Any formal construction must 
\begin{inparaenum}[]
\item detect the states with a deadlock, a lifelock, or a blocking situation,
\item determine the cause of that problem, and
\item make a decision on how to fix the problem.
\end{inparaenum}

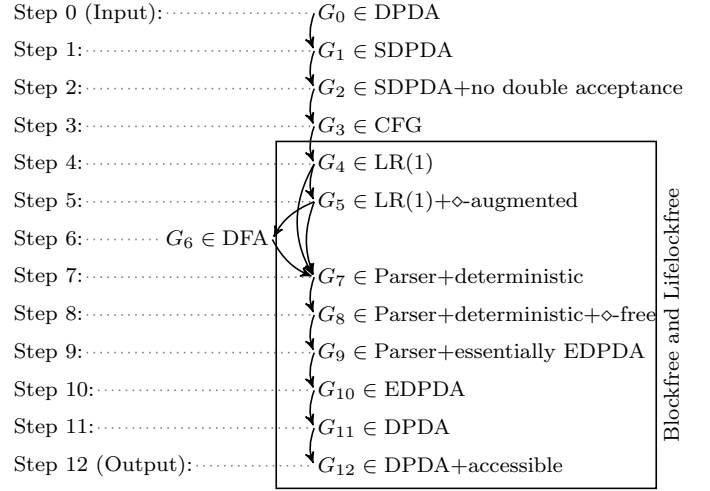
\myFigure{fig__overview}{Visualization of the algorithm.}

\emph{Solution:} In \FIGREF{fig__overview} we have depicted our approach in the subsequently explained 12 steps.
The basic idea is to
\begin{inparaenum}[]
\item (Steps 1--3) transform the \DPDA \STEP{0} into a \CFG \STEP{3},
\item (Step 4) obtain an \LR grammar \STEP{4} by restricting \STEP{3} to establish operational blockfreeness and absence of lifelocks,
\item (Steps 5--11) transform the \LR grammar into a \DPDA \STEP{11} preserving the desired properties, and finally
\item (Step 12) remove all inaccessible states and edges.
\end{inparaenum}

Steps~1--4 and~7--12 preserve the marked language.
Steps~1--3 and~7--12 preserve the unmarked language while step~4 restricts the unmarked language to the prefix closure of the marked language.
Steps~5 and~6 are not meant to preserve the (un)marked language as they are only intermediate results of the translation in Step~7.

\myparagraph{Approximating Accessibility}\label{mypar:ApproximatingAccessibility}
Throughout the following presentation we omit states and edges which are obviously inaccessible: such states and edges are detected by overapproximating the possible ${\leq}k$-length prefixes of stacks in reachable configurations.
The $k$-overapproximation $\mathcal{R}:Q\fun Q\fun \POWERSET{\Gamma^{\leq k}}$ is the least function satisfying the following rules:
\begin{inparaenum}[(i)]
\item initial configuration: $\kPREFIX{\ENDOFSTACK}\in\mathcal{R}(\STATEELEM_0,\STATEELEM_0)$,
\item closure under steps: if $\STACKELEM\STACKSTR\in\mathcal{R}(\STATEELEM,\STATEELEM)$ and $(\STATEELEM,\OPTSIGMAELEM,\STACKELEM,\STACKSTR',\STATEELEM')\in\delta$ then $\kPREFIX{(\STACKSTR'\sconc\STACKSTR)}\in\mathcal{R}(\STATEELEM,\STATEELEM')$ and $\kPREFIX{(\STACKSTR'\sconc\STACKSTR)}\in\mathcal{R}(\STATEELEM',\STATEELEM')$, and
\item transitivity: if $\STACKSTR\in\mathcal{R}(\STATEELEM,\STATEELEM')$ and $\STACKSTR'\in\mathcal{R}(\STATEELEM',\STATEELEM'')$ then $\STACKSTR'\in\mathcal{R}(\STATEELEM,\STATEELEM'')$.\footnote{
Without using the transitivity rule we obtain the $0$- and $1$-overapproximations $\mathcal{R}_0$ and $\mathcal{R}_1$ of \STEP{0} (where we omit empty sets):

$\mathcal{R}_0=
\{
\STATEELEM_0\mapsto\{\STATEELEM_1\mapsto\{\EMPTYSTRING\}, \STATEELEM_0\mapsto\{\EMPTYSTRING\}\},
\STATEELEM_1\mapsto\{\STATEELEM_2\mapsto\{\EMPTYSTRING\}, \STATEELEM_1\mapsto\{\EMPTYSTRING\}, \STATEELEM_3\mapsto\{\EMPTYSTRING\}\},
\STATEELEM_2\mapsto\{\STATEELEM_2\mapsto\{\EMPTYSTRING\}, \STATEELEM_1\mapsto\{\EMPTYSTRING\}\},
\STATEELEM_3\mapsto\{\STATEELEM_1\mapsto\{\EMPTYSTRING\}, \STATEELEM_4\mapsto\{\EMPTYSTRING\}, \STATEELEM_3\mapsto\{\EMPTYSTRING\}\},
\color{red}\mathbf{\STATEELEM_4\mapsto\{\STATEELEM_4\mapsto\{\EMPTYSTRING\}\}}\color{black}
\}
$

$\mathcal{R}_1=
\{
\STATEELEM0\mapsto\{\STATEELEM_1\mapsto\{\ENDOFSTACK, \bullet\}, \STATEELEM0\mapsto\{\ENDOFSTACK, \bullet\}\},
\STATEELEM_1\mapsto\{\STATEELEM_2\mapsto\{\EMPTYSTRING\}, \STATEELEM_1\mapsto\{\EMPTYSTRING, \ENDOFSTACK, \bullet\}, \STATEELEM_3\mapsto\{\bullet\}\},
\STATEELEM_2\mapsto\{\STATEELEM_2\mapsto\{\EMPTYSTRING, \ENDOFSTACK\}, \STATEELEM_1\mapsto\{\EMPTYSTRING\}\},
\STATEELEM_3\mapsto\{\STATEELEM_1\mapsto\{\EMPTYSTRING\}, \STATEELEM_3\mapsto\{\bullet\}\}
\}
$
}
\end{inparaenum}

For example, in $\STEP{0}$ the state $\STATEELEM_4$ is obviously inaccessible because the set of all ${\leq}1$-length prefixes of stacks of reachable configurations with state $\STATEELEM_4$ is empty. However, we would obtain $\EMPTYSTRING$ to be a ${\leq}0$-length prefix of a reachable configuration with state $\STATEELEM_4$; i.e., by increasing the parameter for the length of the calculated prefixes a better result may be obtained. For \DFA and $k=0$ the standard \DFA-accessibility-operation is obtained.
For arbitrary \DPDA step~12 alone enforces accessibility.

\newcommand{\OPNAME}[1]{\ensuremath{f_{\mathrm{#1}}}\xspace}
\newcommand{\FUNtoSDPDA}{\OPNAME{2SDPDA}}
\newcommand{\FUNfreshSymbol}{\OPNAME{fresh}}
\newcommand{\FUNRNoOp}{\OPNAME{NoNoOp}}
\newcommand{\FUNSPP}{\OPNAME{SPP}}
\newcommand{\FUNRMP}{\OPNAME{RMP}}
\newcommand{\FUNSR}{\OPNAME{SR}}
\newcommand{\epdagamma}{\mathrm{epda\underbar{ }gamma}}
\newcommand{\mathlet}{\mathrm{let}~}
\newcommand{\mathin}{~\mathrm{in}~}

Applying this approximation implicitly in the running example, we now describe the steps of the algorithm solving the problem stated in \DEFREF{def:ATS:problem}.

\myparagraph{Step 1}
We transform the \DPDA into a simple \DPDA (called \SDPDA subsequently) such that every edge is of one of three forms: a generating edge $(\STATEELEM,\SIGMAELEM,\STACKELEM,\STACKELEM,\STATEELEM')$, a pop edge $(\STATEELEM,\EMPTYSTRING,\STACKELEM,\EMPTYSTRING,\STATEELEM')$, or a push edge $(\STATEELEM,\EMPTYSTRING,\STACKELEM,\STACKELEM'\STACKELEM,\STATEELEM')$.
The operation consists of four steps:
\begin{inparaenum}[(i)]
\item split every edge of the form $(\STATEELEM,\SIGMAELEM,\STACKELEM,\STACKSTR,\STATEELEM')$ into $(\STATEELEM,\SIGMAELEM,\STACKELEM,\STACKELEM,\STATEELEM'')$ and $(\STATEELEM'',\EMPTYSTRING,\STACKELEM,\STACKSTR,\STATEELEM')$,
\item split every neutral edge of the form $(\STATEELEM,\EMPTYSTRING,\STACKELEM,\STACKELEM,\STATEELEM')$ into $(\STATEELEM,\EMPTYSTRING,\STACKELEM,\circ\STACKELEM,\STATEELEM'')$ and $(\STATEELEM'',\EMPTYSTRING,\circ,\EMPTYSTRING,\STATEELEM')$ for a unique fresh stack symbol $\circ\in\Gamma$,
\item split every rule of the form $(\STATEELEM,\EMPTYSTRING,\STACKELEM,\STACKSTR\STACKELEM',\STATEELEM')$ with $\STACKELEM\neq\STACKELEM'$ into 
$(\STATEELEM,\EMPTYSTRING,\STACKELEM,\EMPTYSTRING,\STATEELEM'')$ and
$(\STATEELEM'',\EMPTYSTRING,\STACKELEM'',\STACKSTR\STACKELEM'\STACKELEM'',\STATEELEM')$ for every $\STACKELEM''\in\Gamma$, and
\item split every rule of the form $(\STATEELEM,\EMPTYSTRING,\STACKELEM,\STACKSTR\STACKELEM,\STATEELEM')$ into $\StrLen{\STACKSTR}$ steps which push a single symbol of $\STACKSTR$ in each step.
\end{inparaenum}
Note that the fresh states to be used in each of the four steps contain the edge for which they have been constructed (i.e., $\STATEELEM''$ in the first step is $(\STATEELEM,\OPTSIGMAELEM,\STACKELEM,\STACKSTR,\STATEELEM')$).
The operation has been adapted from \cite{DBLP:journals/iandc/Knuth65} by 
\begin{inparaenum}
\item correcting the handling of neutral edges involving the \ENDOFSTACK symbol (for example, the self loop at $\STATEELEM_2$ in $\STEP{0}$ would have been handled incorrectly), and by
\item logging the involved edges in the fresh states as explained before.
\end{inparaenum}
\myFigure{fig__example__SDPDA_NDA_LR1_no_rename}{The simple \DPDA \STEP{1}, the simple \DPDA \STEP{2} not exhibiting double marking, and the \LR-grammar \STEP{4}.}
For the \DPDA \STEP{0} from \FIGREF{fig__example__M_0_and_M_I} the \SDPDA \STEP{1} in \FIGREF{fig__example__SDPDA_NDA_LR1_no_rename} results (up to renaming of the states).

\myparagraph{Step 2}
We transform the \SDPDA \STEP{1} into an \SDPDA \STEP{2} such that once the \SDPDA \STEP{2} has generated an output, it has to generate another symbol before entering a marking state again.
For the example automaton \STEP{1} this means that the lifelock at $\STATEELEM_2,\STATEELEM_3$ is problematic.
We are reusing the construction from \cite{DBLP:journals/iandc/Knuth65}:
Every state is duplicated (the duplicated states are neither initial nor marking).
Then, the edges are defined such that the automaton \STEP{2} operates on the original states until it reaches a marking state.
Once this happens, the automaton either remains in the original states by using a generating edge or it switches to the duplicated states.
The automaton remains in the duplicated states until switching to the original states using any generating edge.
For the \SDPDA \STEP{1} from \FIGREF{fig__example__SDPDA_NDA_LR1_no_rename} the \SDPDA \STEP{2} in the same figure results.
Note, the lifelock in $\STATEELEM_1,\STATEELEM_3$ has been removed by the cost of another lifelock in $\overline \STATEELEM_1,\overline \STATEELEM_3$ generating the same output $b$.

\myparagraph{Step 3 \& Step 4} 
We transform the \SDPDA \STEP{2} in step~3 into the \CFG \STEP{3} using a construction from \cite{DBLP:journals/iandc/Knuth65}.
We restrict the \CFG \STEP{3} in step~4 to the \LR grammar \STEP{4} (see \FIGREF{fig__example__SDPDA_NDA_LR1_no_rename}) by removing all productions from \STEP{3} which do not appear in any marking derivation of \STEP{3}.
That is, the accessible and coaccessible part is constructed using a fixed-point algorithm in each case.
For the accessible part: the \emph{accessible} nonterminals are the least set of nonterminals $\mathcal{A}$ such that the initial nonterminal is contained in $\mathcal{A}$ and for any production $\NONTERMELEMA\fun \CFGMIXSTR$: if $\NONTERMELEMA\in\mathcal{A}$, then the nonterminals of $\CFGMIXSTR$ are contained in $\mathcal{A}$.
For the coaccessible part: the \emph{coaccessible} nonterminals are the least set of nonterminals $\mathcal{A}$ such that for any production $\NONTERMELEMA\fun \CFGMIXSTR$: if the nonterminals of $\CFGMIXSTR$ are contained in $\mathcal{A}$ then $\NONTERMELEMA\in\mathcal{A}$.
The equivalence of \STEP{2} and \STEP{4} w.r.t. the marked language can best be understood by comparing the derivations in \FIGREF{fig__example__derivation_SDPDA_LR1G}.
The following three properties explain the correctness of the construction:
\begin{inparaenum}[(i)]
\item The nonterminals of the form $L_{\STATEELEM,\NONTERMELEMA}$ (for example $L_{\STATEELEM_1,\Box}$) guarantee a marking derivation of the \SDPDA starting in $\STATEELEM$ not modifying the stack starting with $\NONTERMELEMA$.
\item The nonterminals of the form $L_{\STATEELEM,\NONTERMELEMA,\STATEELEM'}$ (for example $L_{\STATEELEM_0,\bullet,\STATEELEM_2}$) guarantee a derivation of the \SDPDA starting in $\STATEELEM$ not modifying the stack starting with $\NONTERMELEMA$ and reaching a configuration in which the $\NONTERMELEMA$ is removed and the state $\STATEELEM'$ is reached.
\item For any configuration $(\STATEELEM_1,\SIGMASTR,\STACKELEM_1\dots\STACKELEM_n)$ there are $\STATEELEM_2\dots \STATEELEM_n$ such that 
$(\STATEELEM,\SIGMASTR,\STACKELEM_1\dots\STACKELEM_n)$ is reachable by \STEP{2} \IFF $\SIGMASTR\sconc L_{\STATEELEM_1,\STACKELEM_1,\STATEELEM_2}\dots L_{\STATEELEM_{n-1},\STACKELEM_{n-1},\STATEELEM_n}L_{\STATEELEM_n,\STACKELEM_n}$ is reachable by \STEP{4}.
\end{inparaenum}

\emph{Once step~4 has been completed, for the given \DPDA a marked language equivalent \CFG has been constructed which is lifelockfree, accessible, and operational blockfree (and by that deadlockfree).}

\myFigure{fig__example__derivation_SDPDA_LR1G}{Corresponding initial derivations of the \SDPDA \STEP{2} and the \LR grammar \STEP{4}.}

\myparagraph{Step 5 \& Step 6 \& Step 7}
In these steps we are following, with some modifications, the constructions in \cite{sippu-2}.

In step~5 we are constructing the \ENDOFOUTPUT-augmented version \STEP{5} of \STEP{4}: A new initial nonterminal $S'$ and the production $S'\fun \ENDOFOUTPUT S\ENDOFOUTPUT$ are added where $S$ is the old nonterminal.
This modification allows for a simpler construction procedure of the \LR-machine and the \LR-parser in steps~6 and~7.

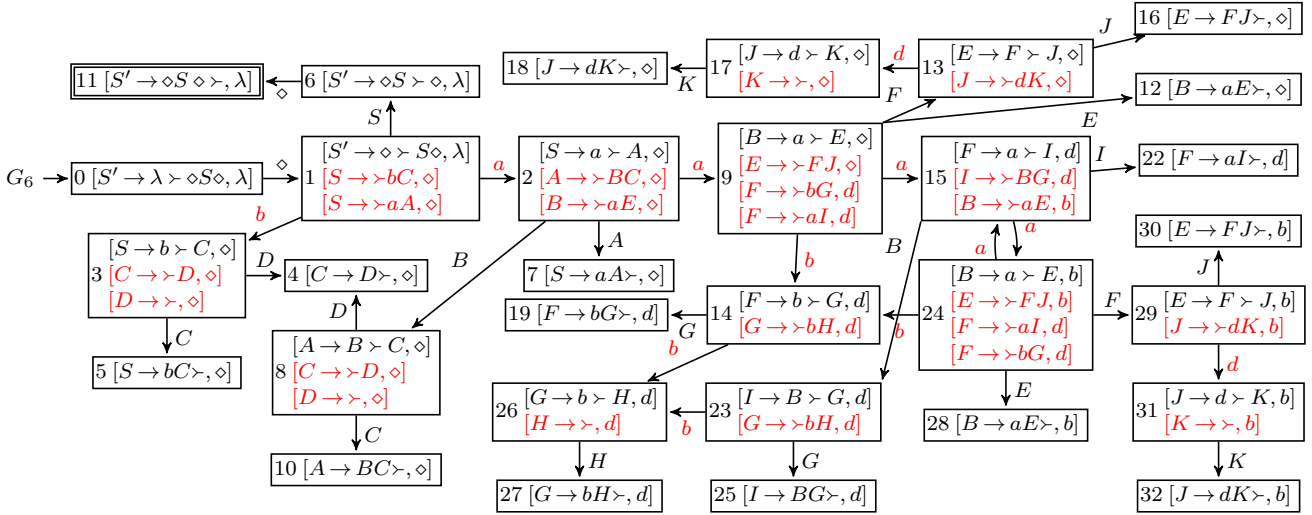
\myFigureWide{fig__example__LRmachine}{The \LR-machine \STEP{6}. Edges generating terminals (relevant for shift-rules) and items with marker $\ITEMMARKER$ at the beginning of the right hand side (relevant for reduce rules) are printed in red.}
In step~6 we are constructing the \LR-machine \STEP{6} (depicted in \FIGREF{fig__example__LRmachine}) for the \LR-grammar \STEP{5}.
The output alphabet of the \DFA \STEP{6} is the union of the output alphabet and the nonterminals of \STEP{5}.
The steps of the parser (between two states $\STATEELEM,\STATEELEM'$ of \STEP{6}) will depend on the elements of $\STATEELEM$: these elements are called items which are formally four-tuples containing a production with a marker splitting the right hand side of the production and a lookahead symbol.
The \DFA \STEP{6} has two kinds of edges:
the edges labeled with an output symbol $\SIGMAELEM$ represent the action where the parser generates $\SIGMAELEM$,
the edges labeled with nonterminals are required for the actions where the parser concludes (based on its stack and the lookahead of the items) that it has generated a word derivable by a nonterminal.

Every edge $(\STATEELEM,\CFGMIXELEM,\STATEELEM')$ in \STEP{6} satisfies that $\STATEELEM'$ is the least set satisfying the following conditions: $\STATEELEM'$ contains all items of $\STATEELEM$ where the marker $\ITEMMARKER$ has been shifted over $\CFGMIXELEM$.
Furthermore, if an item of the form $\ITEM{\NONTERMELEMA}{\CFGMIXSTR}{\NONTERMELEMB\sconc \CFGMIXSTR'}{\OPTSIGMAELEM}$ is obtained, then the so-called \enquote{first}-symbols $\OPTSIGMAELEM'$ are determined\footnote{While in \cite{sippu-2} no effective algorithm is presented for this operation we have been able to verify such a construction.} for which there is a $\SIGMASTR$ satisfying $\CFGMIXSTR'\sconc\OPTSIGMAELEM\DERIVE[*]{\STEP{5}}\SIGMASTR$ with $\OPTSIGMAELEM'=\kPREFIX[1]{\SIGMASTR}$ and for all such (possibly empty) $\OPTSIGMAELEM'$ and all productions of the form $\NONTERMELEMB\fun\CFGMIXSTR''$, the item $\ITEM{\NONTERMELEMB}{}{\CFGMIXSTR''}{\OPTSIGMAELEM'}$ is contained in $\STATEELEM'$\footnote{The state with no items has been removed from the visualization in \FIGREF{fig__example__LRmachine}.}.

For example (in \FIGREF{fig__example__LRmachine}), the $a$-successor of state $9$ is state $15$: $\ITEM{F}{a}{I}{d}\in 15$ is the result of the shifting of the $\ITEMMARKER$ over the $a$ in the item $\ITEM{F}{}{aI}{d}\in 9$; 
$\ITEM{I}{}{BG}{d}\in 15$ because $\ITEM{F}{a}{I}{d}\in 15$ and $d$ is (trivially) derivable to $d$;
$\ITEM{B}{}{aE}{b}\in 15$ because $\ITEM{I}{}{BG}{d}\in 15$ and $Gd$ is derivable to $bd$.

\myFigure{fig__example__Parser0}{The rules of the \LR-parser \STEP{7} with initial state $1$ and marking set $\Set{6}$.}
In step~7 we are constructing the \LR-parser \STEP{7} (depicted in \FIGREF{fig__example__Parser0}) for \STEP{5} and \STEP{6}. The parser consists of shift rules (generating a symbol and changing the stack and state) and reduce rules (which only modify the stack and state). The shift rules are obtained from the \LR-machine by selecting the edges in \STEP{6} which are labeled with an output symbol: an edge $(\STATEELEM,\SIGMAELEM,\STATEELEM')$ would result in the shift rule $\STATEELEM\vert\SIGMAELEM\fun \STATEELEM\sconc \STATEELEM'\vert\EMPTYSTRING$ (e.g., the edge $(1,a,2)$ results in the rule $1\vert a\fun 1\sconc 2\vert\EMPTYSTRING$).
The reduce rules are constructed for every item of the form $\ITEM{\NONTERMELEMA}{}{\CFGMIXSTR}{\SIGMAELEM}\in\STATEELEM$ (i.e., the marker $\ITEMMARKER$ is at the beginning of the right hand side): let $\STATESTR$ (by construction $\STATESTR$ is also a word over the stack alphabet of $\STEP{7}$) be the sequence of states visited by generating $\CFGMIXSTR$ starting in $\STATEELEM$ in \STEP{6} and let $\STATEELEM'$ be the state reached by generating $\NONTERMELEMA$ in $\STATEELEM$ in \STEP{6}.
Then the reduce rule $\STATEELEM\sconc\STATESTR\vert\EMPTYSTRING\fun \STATEELEM\sconc \STATEELEM'\vert\EMPTYSTRING$ is added to 
the parser (e.g., the item $\ITEM{J}{}{dK}{b}\in 29$ results in the rule $29\sconc31\sconc32\vert b\fun 29\sconc30\vert b$).

\begin{remark}
According to \cite{sippu-2}, the parser \STEP{7} is a \emph{correct prefix parser}. However, that is a too weak assertion: their definition of the unmarked language considers a symbol the parser has fixed but not generated not to be part of the generated unmarked word. Since the mode of operation we are interested (control of (embedded) discrete event systems), we had to find new proofs to verify that our stronger condition is also satisfied by the generated parser \STEP{7}.
\end{remark}

\myFigure{fig__example__Parser1}{The rules of the \LR-parser \STEP{8} with initial state $1$ and marking set $\Set{3,17}$ (the nonterminals $\Set{4,5,7,8,10,12,16,18}$ are no longer reachable)}
\myparagraph{Step 8}
Since \DPDA are not capable of terminating the generation by fixing an end-of-output marker, we are modifying the parser \STEP{7} by removing all rules involving the end-of-output marker $\ENDOFOUTPUT$ and by changing the set of marking states such that \STEP{8} (depicted in \FIGREF{fig__example__Parser1}) marks in $(\STACKSTR\sconc\STATEELEM,\SIGMASTR,\FIXEDSTR)$ \IFF some edge $\STACKSTR'\sconc\STATEELEM\vert \ENDOFOUTPUT\fun \STACKSTR''\vert \ENDOFOUTPUT$ has been removed. While it is not mentioned in \cite{sippu-2}, we discovered that this drastic removal of rules preserves the (un)marked language because the parser reaches a configuration in which such an edge is enabled if and only if the stack can be entirely reduced by subsequently executed reduce rules. This optimization also speeds up the parsing process using the presented construction in any other context (e.g., parsing of programming languages for which it has originally been designed).

\myFigure{fig__example__Parser2}{The rules of the \LR-parser \STEP{9} with initial state $1$ and marking set $\Set{(3,\EMPTYSTRING),(17,\EMPTYSTRING)}$.}
\myparagraph{Step 9}
Since \DPDA are not capable of fixing output symbols without generating them, we add the fixed output component of a configuration into the state of the configuration.
For every shift rule of the form $\RULE{\STATEELEM}{\SIGMAELEM}{\STATEELEM\sconc\STATEELEM'}{\EMPTYSTRING}$ the rules $\RULE{(\STATEELEM,\EMPTYSTRING)}{\SIGMAELEM}{\STATEELEM\sconc(\STATEELEM',\EMPTYSTRING)}{\EMPTYSTRING}$ and $\RULE{(\STATEELEM,\SIGMAELEM)}{\EMPTYSTRING}{\STATEELEM\sconc(\STATEELEM',\EMPTYSTRING)}{\EMPTYSTRING}$ are used.
For every reduce rule of the form $\RULE{\STACKSTR\sconc\STATEELEM}{\SIGMAELEM}{\STACKSTR'\sconc\STATEELEM'}{\SIGMAELEM}$ the rules $\RULE{\STACKSTR\sconc(\STATEELEM,\EMPTYSTRING)}{\SIGMAELEM}{\STACKSTR'\sconc(\STATEELEM',\SIGMAELEM)}{\EMPTYSTRING}$ and $\RULE{\STACKSTR\sconc(\STATEELEM,\SIGMAELEM)}{\EMPTYSTRING}{\STACKSTR'\sconc(\STATEELEM',\SIGMAELEM)}{\EMPTYSTRING}$ are used.
The resulting parser \STEP{9} is depicted in \FIGREF{fig__example__Parser2}.

\emph{It is then possible to verify, that all reachable configurations of the resulting parser \STEP{9} have an empty fixed output component.
We call the parser \STEP{9} essentially \EDPDA because it uses none of the extra capabilities of the parser formalism.}

\myparagraph{Step 10}
The essentially \EDPDA parser \STEP{9} can be translated into the \EDPDA \STEP{10} (depicted in \FIGREF{fig__example__EDPDA}) by using for every rule of the form $\RULE{\STACKSTR\sconc\STATEELEM}{\OPTSIGMAELEM}{\STACKSTR'\sconc\STATEELEM'}{\EMPTYSTRING}$ the edge $(\STATEELEM,\OPTSIGMAELEM,\STACKSTR^{-1},\STACKSTR'^{-1},\STATEELEM')$. Marking and initial states of \STEP{10} are taken from \STEP{9}.

\myparagraph{Step 11}
Since \DPDA are not capable of popping strictly \emph{more} than one symbol from the stack, we split such edges into multiple edges to obtain the \DPDA \STEP{11}. To preserve determinism, the splitting of edges with the same source entails the merging of partially identical edges until the recursive split identifies their distinctness. For example, the edges $(\STATEELEM,\OPTSIGMAELEM,\STACKSTR\sconc\STACKSTR',\STACKSTR_1,\STATEELEM_1)$ and $(\STATEELEM,\OPTSIGMAELEM,\STACKSTR\sconc\STACKSTR'',\STACKSTR_2,\STATEELEM_2)$ share a common prefix $\STACKSTR$ on the popping component.

Since \DPDA are not capable of popping strictly \emph{less} than one symbol from the stack, we modify the automaton by replacing any edge $(\STATEELEM,\OPTSIGMAELEM,\EMPTYSTRING,\STACKSTR,\STATEELEM')$ with $(\STATEELEM,\OPTSIGMAELEM,\STACKELEM,\STACKSTR\sconc\STACKELEM,\STATEELEM)$ for any $\STACKELEM$ of the stack alphabet of \STEP{10}. For soundness, recall that the stack-bottom-marker can never be removed from the stack.

\myparagraph{Step 12}
Finally, accessibility of states and edges can be enforced by reusing the presented steps~1--4.
For a \DPDA we are executing steps~1--4.
From the productions obtained by step~4 we can determine by executing the steps~1--3 backwards (which are by our construction injective in the sense that for each constructed production/edge $x$ a unique edge $e$ can be determined for which $x$ has been constructed).
Using this backwards computation, we are able to determine the accessible edges of a \DPDA.
The accessible states are the sources and edges of any of the accessible edges.
The inaccessible states and edges are then removed to obtain the \DPDA \STEP{12} from \FIGREF{fig__example__DPDA}.

We are not aware of comparable constructions ensuring accessibility of \DPDA, however, using the decidability of emptiness from \cite{HopcroftUllman} it is possible to test a single (and by that every) edge for accessibility; this approach has been used in~\cite{Griffin2006}.
Our approach is superior as we are executing a single test on all edges simultaneously.

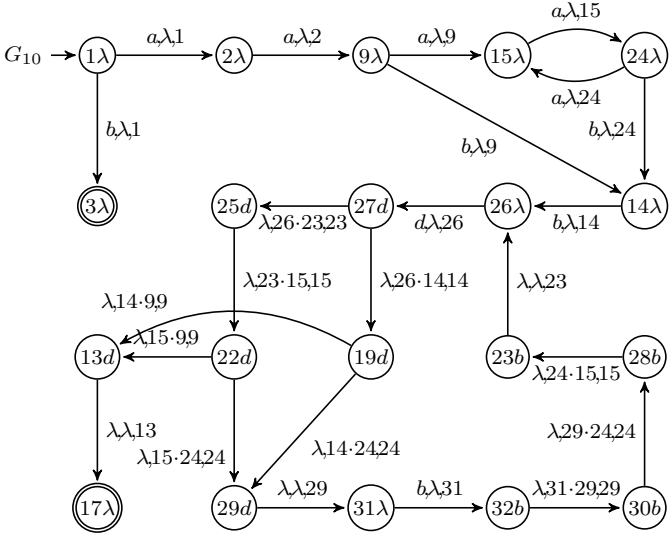
\myFigure{fig__example__EDPDA}{The resulting \EDPDA \STEP{10} where obviously unreachable states have been removed.}
\myFigure{fig__example__DPDA}{The resulting \DPDA \STEP{12}.}

\myparagraph{Verification}
The soundness of the presented algorithm (w.r.t. the problem Definition~\ref{def:ATS:problem}) has been verified in the interactive theorem prover Isabelle/HOL~\citep{IsabelleHOL2011} apart from the following steps for which only pen-and-paper proofs exist yet and which are to be completed in Isabelle/HOL in the near future: 
\begin{inparaenum}[(i)]
\item the \CFG obtained in step~4 is an \LR grammar (satisfied according to \cite{DBLP:journals/iandc/Knuth65}),
\item the \Parser obtained in step~7 is deterministic if \STEP{5} is an \LR grammar (satisfied according to \cite{sippu-2}),
\item step~11, and
\item step~12.
\end{inparaenum}
From these tasks however, only the first appears to be complicated.

\myparagraph{Testing}
The presented algorithm has been implemented in Java for rapid prototyping and in C++ as a plugin to the \cite{libFAUDES} tool.
The implementations have been used successfully for many examples including the running example of this paper.

\myparagraph{Optimizations}
The algorithm can be optimized in different ways.
\begin{inparaenum}[(i)]
\item The runtime of the algorithm depends primarily on the steps~3 and~4 because \STEP{3} would have an enormous amount of productions. We can greatly restrict the set of productions to be generated by exploiting the structure of the input \DPDA using the reachability overapproximation presented on page~\pageref{mypar:ApproximatingAccessibility}.
\item Furthermore, steps~3 and~4 can be merged such that only productions are generated which are coaccessible.
This alternative trades runtime for space-requirements (the size of \STEP{4} is usually not much greater than \STEP{1} but the runtime is increased by the length of the longest derivation necessary in \STEP{2} to reach all states).
\item Another optimization merges adjacent edges in \EDPDA which are intermediate results. This optimization decreases the runtime of the subsequently executed operations.
\end{inparaenum}
The formal definition and verification in Isabelle/HOL of such intermediate operations is left for future work.

\section{Conclusion}\label{sect:SUM}
The algorithm presented in this paper optimizes the behavior of a \DPDA whilst preserving its marked language by first translating the \DPDA into another model (\LR grammars) in which the desired properties can be enforced using simple constructions and by translating the obtained solution back into \DPDA while preserving the desired properties.

The algorithm guarantees accessibility (every state and every edge is required for some marking derivation), lifelockfreeness (there is no initial derivation executing infinitely many steps without generating an output symbol), deadlockfreeness (non-extendable initial derivations are ending in marking states), and finally the operational blockfreeness (every initial derivation can be extended into a marking derivation).

The operational blockfreeness is sufficient to conclude that the unmarked language is the prefix closure of the marked language of the resulting \DPDA.

The algorithm does not minimize the size of the automaton, in fact, the size of the resulting \DPDA is usually increased and is growing according to \cite{DBLP:conf/focs/GellerHSU75} in some cases exponentially.

The algorithm presented here is a crucial part of the presented solution of the supervisory control problem for \DFA plants and \DPDA specifications which is reduced (in the companion paper by \citet*{SchneiderSchmuck2014}) to the effective implementability of ensuring blockfreeness (solved in this paper) and ensuring controllability (solved in the companion paper by \citet*{SchmuckSchneider2014}).

\section{Future Work}\label{sect:FUTURE}
\myparagraph{Petri nets}
Since the problem of establishing blockfreeness is unsolvable for standard Petri net classes~\citep{GiuaCesare1994,GiuaCesare1995}, we intend to determine Petri net classes $\mathcal{P}$ that can be translated (preserving the marked language) into a \DPDA $G$ such that the \DPDA generated by our algorithm $G'$ can be translated back into a Petri net from $\mathcal{P}$ to solve the problem for such a Petri net class.

\myparagraph{Visibly Pushdown Tree Automata (\VPTA)}
\VPTA introduced by \cite{DBLP:conf/frocos/ChabinR07} are the greatest known subclass of \DPDA which are closed under intersection.
For the context of the Supervisory Control Theory we intend to determine an algorithm which solves the problem from \DEFREF{def:ATS:problem} for \VPTA because
\begin{inparaenum}[(i)]
\item plant and controller can then be generated by \VPTA, while this decreases the expressiveness for the controller language it also increases the expressiveness for the plant language, and
\item the closed loop is again a \VPTA, which allows for the iterative restriction of a plant language by horizontal composition of controllers.
\end{inparaenum}
The algorithm presented here may be reusable: the output of the algorithm, when executed on a \VPTA, may be (convertible) into a \VPTA.
Therefore, when using \VPTA for plants, specifications, and controllers, the supervisory controller synthesis can be extended to yet another domain.

\myparagraph{Nondeterminism}
For the context of the Supervisory Control Theory there is no reason to restrict oneself to deterministic controllers. However, for these systems the desired property of operational blockfreeness is not guaranteed for language blockfree controllers. Therefore, when extending the domain of the algorithm to \PDA the proofs will become more complex as the preservation of marked and unmarked language is no longer sufficient for the preservation of the operational blockfreeness as discussed in \cite{SchneiderSchmuck2014}.


\begin{thebibliography}{15}
\providecommand{\natexlab}[1]{#1}
\providecommand{\url}[1]{\texttt{#1}}
\providecommand{\urlprefix}{URL }
\expandafter\ifx\csname urlstyle\endcsname\relax
  \providecommand{\doi}[1]{doi:\discretionary{}{}{}#1}\else
  \providecommand{\doi}{doi:\discretionary{}{}{}\begingroup
  \urlstyle{rm}\Url}\fi

\bibitem[{Aho and Ullman(1972)}]{Aho:1972:TPT:578789}
Aho, A.V. and Ullman, J.D. (1972).
\newblock \emph{The theory of parsing, translation, and compiling}.
\newblock Prentice-Hall, Inc., Upper Saddle River, NJ, USA.

\bibitem[{Chabin and R{\'e}ty(2007)}]{DBLP:conf/frocos/ChabinR07}
Chabin, J. and R{\'e}ty, P. (2007).
\newblock Visibly pushdown languages and term rewriting.
\newblock In B.~Konev and F.~Wolter (eds.), \emph{FroCoS}, volume 4720 of
  \emph{Lecture Notes in Computer Science}, 252--266. Springer.

\bibitem[{Geller et~al.(1975)Geller, III, Szymanski, and
  Ullman}]{DBLP:conf/focs/GellerHSU75}
Geller, M.M., III, H.B.H., Szymanski, T.G., and Ullman, J.D. (1975).
\newblock Economy of descriptions by parsers, dpda's, and pda's.
\newblock In \emph{FOCS}, 122--127. IEEE Computer Society.

\bibitem[{Ginsburg and Greibach(1966)}]{DBLP:journals/iandc/GinsburgG66b}
Ginsburg, S. and Greibach, S.A. (1966).
\newblock Deterministic context free languages.
\newblock \emph{Information and Control}, 9(6), 620--648.

\bibitem[{Giua and DiCesare(1994)}]{GiuaCesare1994}
Giua, A. and DiCesare, F. (1994).
\newblock Blocking and controllability of petri nets in supervisory control.
\newblock \emph{IEEE Transactions on Automatic Control}, 39(4), 818--823.
\newblock \doi{10.1109/9.286260}.

\bibitem[{Giua and DiCesare(1995)}]{GiuaCesare1995}
Giua, A. and DiCesare, F. (1995).
\newblock Decidability and closure properties of weak petri net languages in
  supervisory control.
\newblock \emph{IEEE Transactions on Automatic Control}, 40(5), 906--910.
\newblock \doi{10.1109/9.384227}.

\bibitem[{Griffin(2006)}]{Griffin2006}
Griffin, C. (2006).
\newblock A note on deciding controllability in pushdown systems.
\newblock \emph{IEEE Transactions on Automatic Control}, 51(2), 334 -- 337.

\bibitem[{Hopcroft and Ullman(1979)}]{HopcroftUllman}
Hopcroft, J.E. and Ullman, J.D. (1979).
\newblock \emph{Introduction to Automata Theory, languages and computation}.
\newblock Addison-Wesley Publishing company.

\bibitem[{Knuth(1965)}]{DBLP:journals/iandc/Knuth65}
Knuth, D.E. (1965).
\newblock On the translation of languages from left to rigth.
\newblock \emph{Information and Control}, 8(6), 607--639.

\bibitem[{libFAUDES(2006-2013)}]{libFAUDES}
libFAUDES (2006-2013).
\newblock Software library for discrete event systems.
\newblock \urlprefix\url{http://www.rt.eei.uni-erlangen.de/FGdes/faudes}.

\bibitem[{Paulson et~al.(2011)Paulson, Nipkow, and Wenzel}]{IsabelleHOL2011}
Paulson, L., Nipkow, T., and Wenzel, M. (2011).
\newblock Isabelle/{HOL}.
\newblock \urlprefix\url{http://isabelle.in.tum.de}.

\bibitem[{Schmuck et~al.(2014)Schmuck, Schneider, Raisch, and
  Nestmann}]{SchmuckSchneider2014}
Schmuck, A.-K., Schneider, S., Raisch, J., and Nestmann, U. (2014).
\newblock Extending supervisory controller synthesis to deterministic pushdown
  automata---enforcing controllability least restrictively.
\newblock \emph{WODES'14}.

\bibitem[{Schneider and Schmuck(2013)}]{SchneiderSchmuck_TechRep_2013}
Schneider, S. and Schmuck, A.-K. (2013).
\newblock Supervisory controller synthesis for deterministic pushdown automata
  specifications.
\newblock Technical report, Technical University of Berlin, \urlprefix\url{http://www.tu-berlin.de/?25631}.

\bibitem[{Schneider et~al.(2014)Schneider, Schmuck, Raisch, and
  Nestmann}]{SchneiderSchmuck2014}
Schneider, S., Schmuck, A.-K., Raisch, J., and Nestmann, U. (2014).
\newblock Reducing an operational supervisory control problem by decomposition
  for deterministic pushdown automata.
\newblock \emph{WODES'14}.

\bibitem[{Sippu and Soisalon-Soininen(1990)}]{sippu-2}
Sippu, S. and Soisalon-Soininen, E. (1990).
\newblock \emph{Parsing Theory}, volume II: {LR}($k$) and {LL}($k$) Parsing of
  \emph{EATCS Monographs on Theoretical Computer Science}.
\newblock Springer-Verlag.

\end{thebibliography}
\end{document}